\documentclass[%
 aps, prapplied,
 amsmath,amssymb,
 reprint,%
 superscriptaddress,
]{revtex4-2}

\usepackage{graphicx}
\usepackage{dcolumn}
\usepackage{bm}
\usepackage{mathbbol}

\usepackage[utf8]{inputenc}
\usepackage[T1]{fontenc}
\usepackage{mathptmx}
\usepackage{comment}
\usepackage{xcolor}

\DeclareMathOperator{\sech}{sech}

\begin{document}


\title[Spin-Hall nanooscillator based on an antiferromagnetic domain wall.]{Spin-Hall nanooscillator based on an antiferromagnetic domain wall.}

\author{R.V. Ovcharov}
\affiliation{ 
Department of Physics, University of Gothenburg, Gothenburg 41296, Sweden
}

\author{E. G. Galkina}
\affiliation{ 
Institute of Physics, National Academy of Science of Ukraine, 03142 Kiev, Ukraine
}

\author{B.A. Ivanov}
\affiliation{
Institute of Magnetism of the National Academy of Sciences of Ukraine and the Ministry of Education and Science of Ukraine, Kiev 03142, Ukraine
}
\author{R.S. Khymyn}%
\affiliation{ 
Department of Physics, University of Gothenburg, Gothenburg 41296, Sweden
}

\date{\today}

\begin{abstract}
	We propose here a high-frequency spin-Hall nano-oscillator based on a simple magnetic texture, such as a domain wall, located in an antiferromagnet with an easy-axis anisotropy type. We show that the spin current, polarized along the anisotropy axis, excites a conical precession of the N\'eel vector in such a domain wall, which allows obtaining a robust ac output signal -- contrary to the planar precession in a uniform uniaxial antiferromagnet, where ac output is hard to achieve. The frequency of the auto-oscillations is easily tunable by the applied current up to the THz range, and the threshold current vanishes for pure uniaxial antiferromagnet. By micro-magnetic simulations, we demonstrate that the pinning of the domain wall is crucial for the oscillator design, which can be achieved in the nano-constriction layout of the free layer. 
\end{abstract}

\maketitle

Spin-transfer-torque and spin-Hall nano-oscillators (STNOs and SHNOs) are well-established devices in modern spintronics \cite{demidov2012magnetic, demidov2014nanoconstriction, slavin2009nonlinear, chen2016spin}. They can act as tiny frequency generators, or as strongly nonlinear ``active'' elements for the advanced signal processing, including neuromorphic \cite{grollier2020neuromorphic} and stochastic \cite{locatelli2014spin} computing. Both types of oscillators consists of a ``free'' magnetic layer and an adjunct spin-current source. Spin torque, which arises from the input spin current, drives the magnetization dynamics, which then can be readout as an output electric alternate current. The operational frequency of ferromagnetic devices is defined by the resonant modes of the magnetic layer, i.e., is determined by the bias magnetic field \cite{bonetti2010experimental, houshang2018spin} and usually lays in 1–50 GHz range \cite{bonetti2009spin}, but in practice rarely exceeds 30 GHz. 

A significant growth of the operating frequencies, even in the absence of an external field, can be realized in portable spintronics devices by employing antiferromagnetic \cite{cheng2016terahertz, khymyn2017antiferromagnetic} (AFM) or compensated ferrimagnetic \cite{ivanov2019ultrafast, lisenkov2019subterahertz} materials for ``free'' layers, where a strong exchange field has a definitive contribution to the spin dynamics \cite{baltz2018antiferromagnetic, ivanov2020spin, gomonay2014spintronics}. Thus, the so-called exchange enhancement is applied to the main AFM dynamic parameters: it leads to the ultra-high frequencies of the magnetic resonance \cite{turov2010symmetry}, which can reach a THz frequency range (0.3-3.0 THz)  and substantially high limiting velocity of domain wall motion \cite{bar2006dynamics}. Exchange enhancement also occurs for non-conservative phenomena, such as magnetic damping and spin-transfer torque. Besides, the AFMs can conduct \cite{wang2014antiferromagnonic}, rectify \cite{khymyn2017antiferromagneticRectifier} and even amplify \cite{wang2014antiferromagnonic, khymyn2016transformation} the spin current.

In the recently proposed AFM-based SHNOs the spin current, polarized along $\textbf{p}$, induces a torque on the  N\'eel vector $\textbf{l}$, which starts to rotate in the plane perpendicular to the $\textbf{p}$ \cite{gomonay2010spin,gomonay2014spintronics,cheng2016terahertz,khymyn2017antiferromagnetic}. The dynamics of the N\'eel vector, in turn, can induce an output electric current 
\begin{equation}
\mathbf{j}_{out} \propto \tau_{out}=[\mathbf{l} \times \mathbf{\dot l}]
\label{jout}
\end{equation}
by spin-pumping and inverse spin-Hall mechanisms. Since for a planar rotation $\mathbf{\dot l} \perp \mathbf{l} \perp \mathbf{p}$ (known as proliferation \cite{cheng2016terahertz}), the alternate output is present only for the nonuniform in time N\'eel vector dynamics. This method is applicable to the bi-axial AFM with the easy-plane type of primary magnetic anisotropy \cite{khymyn2017antiferromagnetic, ivanov2020spin}, where the in-plane potential created by the small secondary anisotropy accelerates/decelerates the rotation of N\'eel vector. This potential, however, creates a threshold for auto-oscillations that would require the application of extremely high currents for its overcoming in the case of easy-axis AFMs. Besides, the above method produces a substantial ac output only not far above the threshold that limits a useful frequency range.

\begin{figure}[hbt!]
\includegraphics[width=\linewidth]{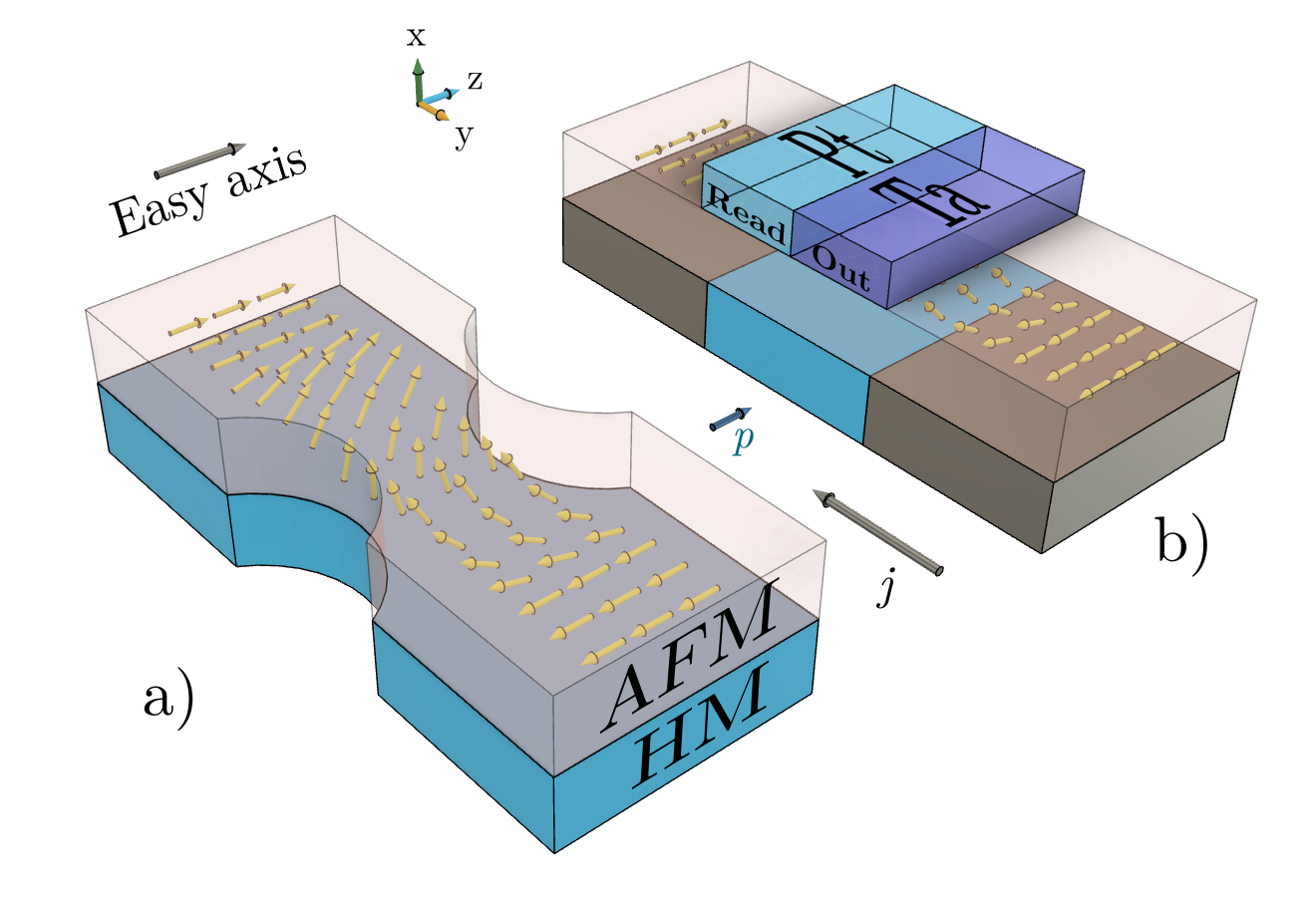}
\caption{\label{fig:schema}  Schematic diagrams illustrating a microwave generator with a thin film of a uniaxial AFM with domain wall used as an active element in a) nano-constriction and b) rectangular geometries. The Neel vector is shown by the yellow arrows, the black arrow shows the electrical current direction, the blue arrow indicates the direction of the spin current polarization. }
\end{figure}%

Another way to get the ac signal is to excite the conical precession of $\mathbf{l}$ for easy-axis AFM \cite{cheng2016terahertz}, or for AFMs in the so-called cone phase \cite{ivanov2020spin}. To achieve this, here we propose to use the natural heterogeneity of magnetic order inside magnetic solitons; various types of those already play a significant role in ferromagnetic-based spintronics \cite{hoefer2010theory, mohseni2013spin, zhou2015dynamically, iacocca2014confined, chung2016magnetic, chung2018direct, sato2019domain}. We consider the simplest topologically stable (the topological charge $\pi_0$) soliton, describing $180^\circ$ domain wall \cite{gomonay2018antiferromagnetic} (DW) and show that nano-oscillators based on AFM DW can create a substantial ac output while keeping the benefits of the AFM materials, such as ultra-high achievable frequency. We are focused on the spin-Hall geometry of the device; however, the developed theory of the DW dynamics is also applicable for STNOs with the $90^\circ$ rotation of the appropriate axes.

The AFM DW motion under the action of spin-orbit torques already aroused considerable interest in the literature \cite{hals2011phenomenology, cheng2014dynamics}. Particularly, it was shown that both non-conservative (damping-like) and conservative (field-like) torques can drive the translational motion of the DW with ultra-high velocities \cite{shiino2016antiferromagnetic, gomonay2016high, li2004domain}, which are two orders of magnitudes higher than the ones in ferromagnets. Here we are focused on the dynamics of the DW in the easy-axis AFM excited by the damping-like torque.   


For our scheme, the possibility of controlling DWs in thin film of AFM is of crucial importance. The presence of a weak non-compensated magnetic moment makes the solution simple for easy-axial canted AFM-like orthoferrites. For DyFeO$_3$ canted (weak ferromagnetic) phase is present above so-called Morin point, approximately 50 K. Thanks to this magnetic moment, a quite-regular domain structure is well known for canted AFMs (see, e.g., Fig. 6 in the review article \cite{bar1985dynamics}). An application of non-uniform in space and weak enough magnetic field with the field gradient as small as 0.1 T/cm stabilizes two-domain configuration with a single DW placed on the line of zero-field (see Fig. 10 in \cite{bar1985dynamics}). Even for “pure” AFMs, in particular, for the antiferromagnetic phase of DyFeO$_3$ below the Morin point, DWs were observed optically and controlled by the usage of a combination of a magnetic field and stress \cite{afanasiev2016control}. The aforementioned results correspond to films with a thickness of the order of microns, but domain structures are known even for ultrathin magnetic films, and even for magnetic monolayers \cite{vedmedenko2000magnetic}. Epitaxial films of dysprosium orthoferrite with thicknesses down to 5 – 10 nm were grown using the pulsed laser deposition \cite{khim2011strain}. For these films, the magnetic structure, including the value of the weak magnetic moment of the order of 0.05 $\mu_B$ on iron ion, is the same as for bulk material. 

The low-energy (in comparison with the energy of the exchange interaction) spin excitation of an AFM can be described by a $\sigma$-model equation with a single variable -- unit N\'eel vector $\mathbf{l}$ \cite{gomonay2014spintronics,ivanov2019ultrafast}, (normalized by the saturated AFM magnetization value $M_s$, which corresponds to the parallel orientation of both sublattices). The vector $\mathbf{l}$ is convenient to represent in angular variables:
\begin{equation}
l_x=\sin\theta \cos\phi, l_y=\sin\theta \sin\phi, l_z=\cos\theta,
\end{equation}
where the axis $z$ is chosen along the easy axis of the AFM, so that the ground state corresponds to $\theta = 0, \pi$. The Lagrangian density is then \cite{kosevich1990magnetic, gomonay2010spin}:

 \begin{equation}
 \mathcal{L}=\frac{M_s}{2\gamma \omega_{ex}}\left[\dot \theta^2 -c^2 \theta'^2+\sin^2 \theta (\dot \phi^2 -c^2 \phi'^2)\right]-w_a,
 \label{lagrangian}
 \end{equation}
where upper dot and prime denote derivatives over time and space respectively, and  $\gamma$ is a gyromagnetic ratio, $\omega_{ex}=\gamma H_{ex}$ is the frequency defined by the uniform exchange field $H_{ex}$ of the AFM, $c=x_0 \omega_0=\gamma\sqrt{H_{ex}A/M_s}$ -- characteristic speed of magnons, which can be written through the domain-wall thickness $x_0=\sqrt{A/K}$ and the magnon gap  $\omega_0=\gamma\sqrt{H_{ex}K/M_s}$, where $A$ is the inhomogeneous exchange constant. The energy density of the anisotropy reads as:
\begin{equation}
w_a=\frac{M_s}{2\gamma \omega_{ex}}(\omega_0^2+\omega_{ip}^2\sin^2\phi)\sin^2\theta,\label{eq:anisotropy}
\end{equation} 
where the first term defines purely uniaxial anisotropy $K$ of the easy-axis type, and the second term defines anisotropy $K_{ip}$ in the basal plane for an AFM with an easy axis of the second order $C_{2}$. We are using a finite value of $K_{ip}$ to compare pure uniaxial and non-uniaxial cases.

The natural dissipation and the influx of energy by STT can be expressed in Rayleigh dissipation function as \cite{ivanov2020spin, gomonay2010spin}:
\begin{eqnarray}
\mathcal{R}=&&\frac{\alpha M_s }{2\gamma}(\dot \theta^2+\dot\phi^2 \sin^2 \theta )+ \nonumber\\
&& \frac{\tau M_s }{\gamma}[p_x(\dot\theta\sin\phi +\dot\phi\cos\phi \sin 2 \theta)-p_z\dot\phi\sin^2 \theta],
\label{eq:rayleigh}
\label{dissip}
\end{eqnarray}
where $\alpha$ is an effective Gilbert damping, $\mathbf{p}$ is the unit vector along with spin current polarization, and $\tau$ is the amplitude of the STT, expressed in the units of frequency $\tau = \sigma j$, $j$ is the density of the electrical current, $\sigma$ -- STT efficiency \cite{slavin2009nonlinear, slavin2006theory}. 

The rotation of the vector $\mathbf{l}$ in the AFM driven by the spin pumping mechanism generates an output spin current (Eq. \ref{jout}) into adjunct layer with \cite{lisenkov2019subterahertz, ivanov2020spin, khymyn2017antiferromagnetic}:
\begin{equation}\label{eq:outputJ}
    \mathbf{\tau}_{out} = \omega \left[ \mathbb{z} \sin^2 \theta - \sin \theta \cos \theta (\mathbb{x} \cos \omega t + \mathbb{y} \sin \omega t)\right],
\end{equation}
From Eq. (\ref{eq:outputJ}) the condition $\sin \theta  \cos \theta \neq 0$ is required to obtain the ac output. However, in the case of uniform spin dynamics, the angle of stationary precession is determined from the condition $d w_a / d \theta = 0$. From Eq. (\ref{eq:anisotropy}),  $d w_a / d \theta \propto \sin \theta \cos \theta$, which corresponds to $\theta = \pi / 2$, i.e. the ac signal is absent. Contrary, an ac output appears in the nonuniform state of the N\'eel order parameter, or spin texture, in the region where $\theta \neq \pi/2$.  The simplest example of such a spin texture is a dynamical domain wall with the known profile \cite{kim2014propulsion,galkina2017precessional, ivanov1995solitons}:
\begin{equation}\label{eq:domainWallProfile}
\cos \theta=\tanh \left(\frac{x-X(t)}{\Delta}\right),  \qquad \phi= \Phi(t),
\end{equation}
where $\Delta$ is an instant value of the DW thickness; $\Delta=x_0$ for the stationary DW.
Using the solution of the DW profile (\ref{eq:domainWallProfile}) and assuming $\alpha \ll 1$ and $\tau \ll \omega_0$, we can write down the equations of motion through the collective coordinates: the coordinate of the wall center -- $X$ and the angle $\Phi$, which determines the rotation angle of the vector $\textbf{l}$ in the wall center (See Supplementary materials):
\begin{eqnarray}
    &&\frac{1}{\omega_{ex}} \frac{d}{dt} \left( \frac{ \dot X }{ \Delta} \right)  +   \alpha \frac{\dot{X}}{\Delta}  +  \tau\frac{\pi }{2}( p_x \sin \Phi - p_y  \cos \Phi) = 0,\label{eq:main} \\
       &&\frac{1}{ \omega_{ex}} \frac{d}{dt} ( \Delta \dot \Phi) + \alpha \Delta \dot \Phi  - \Delta\tau p_z   + \Delta\frac{\omega_{ip}^2}{2\omega_{ex}} \sin 2\Phi = 0,\label{eq:main2}
\end{eqnarray}
where  $\Delta = x_0 \left. \sqrt{1 - \dot X^2/c^2} \middle/\sqrt{1 - \dot \Phi^2/\omega_0^2}\right.$.

The equations (\ref{eq:main}, \ref{eq:main2}) are one of the main analytical results of this work: they show the possibility of excitation of the both rotational ($p_z \neq 0$) and translational ($p_x, p_y \neq 0$) dynamics of the domain wall by the spin current. With this in mind, below, we discuss in detail the possibility of creating a spin-torque nanogenerator based on the AFM domain wall and analyze the regimes of its operation depending on the polarization direction of the spin current.  

To verify analytical results, we performed micro-magnetic simulations using \emph{MuMax3} solver \cite{vansteenkiste2014design} employing method similar to the described in  Ref. \cite{de2017modelling}. We chose two geometries of the SHNO: the rectangular one, which completely represents our analytical model, and the nano-constriction (NC, See Fig. \ref{fig:schema}). The NC cut-out not only increases the local current density but also induces pinning potential for the DW since the minimal length and accordingly the minimal energy of the DW is reached in the center of the NC. Therefore, in a pure AFM without another induced pinning center, a DW created anywhere within an NC area will naturally relax into the center position, which makes this geometry advantageous for practical applications. As an AFM layer, we assumed dysprosium orthoferrite DyFeO$_3$, in which the anisotropy in the easy plane changes its sign at the temperature $T=150$ K, and hence can be chosen arbitrary weak (this property has been established by investigation of DW structure \cite{zalesskij1975nmr}, magnetic resonance measurements \cite{balbashov1985high} and pump-probe technique \cite{kimel2005ultrafast}). As a spin-Hall layer, Pt is chosen. Thus, we chose the following parameters \cite{nguyen2016spin, turov2010symmetry}: $\theta_{SH} = 0.1$, $\alpha=10^{-3}$, $M_s = 8.4 \cdot 10^5$ A/m, $A=18.9$pJ/m, $H_{ex}=670$T, the anisotropy constant along the easy axis $K = 300$ kJ/m$^3$, the value of the weak secondary AFM anisotropy $K_{ip} = 2$ kJ/m$^3$.  
Simulations were performed for the 136 nm wide sample (for both cases) with a centrally located NC with a width of 100 nm and a cut-out radius of $50$ nm. For the rectangular shape, the length of the Pt layer was limited to $L=100$ nm to study the stability of the DW position.

For the further analysis of the DW dynamics, we introduce the angle $\psi$, which determines the direction of the polarization vector $\mathbf{p}=\mathbb{z}\cos\psi+\mathbb{y}\sin\psi$. The angle  $\psi$ is easily configurable in STNOs, where the adjunct layer determines the polarization of the spin current. However, in SHNOs $\mathbf{p}$ is strictly determined by the direction of the electrical current, and in the chosen configuration corresponds to $z$- direction. Thus, the direction of the easy axis should be changed to obtain a non-zero angle $\psi$. In this case, the domain wall itself does not change its position, as it is not bound to the spin frame of reference, but the direction of the output torque is changed, see details below. For this case, we introduce another angle $\varphi_{DW}$ which determines the deviation of the easy axis relative to the DW orientation ($z$-axis).

\begin{figure}[t!]
\includegraphics[width=\linewidth]{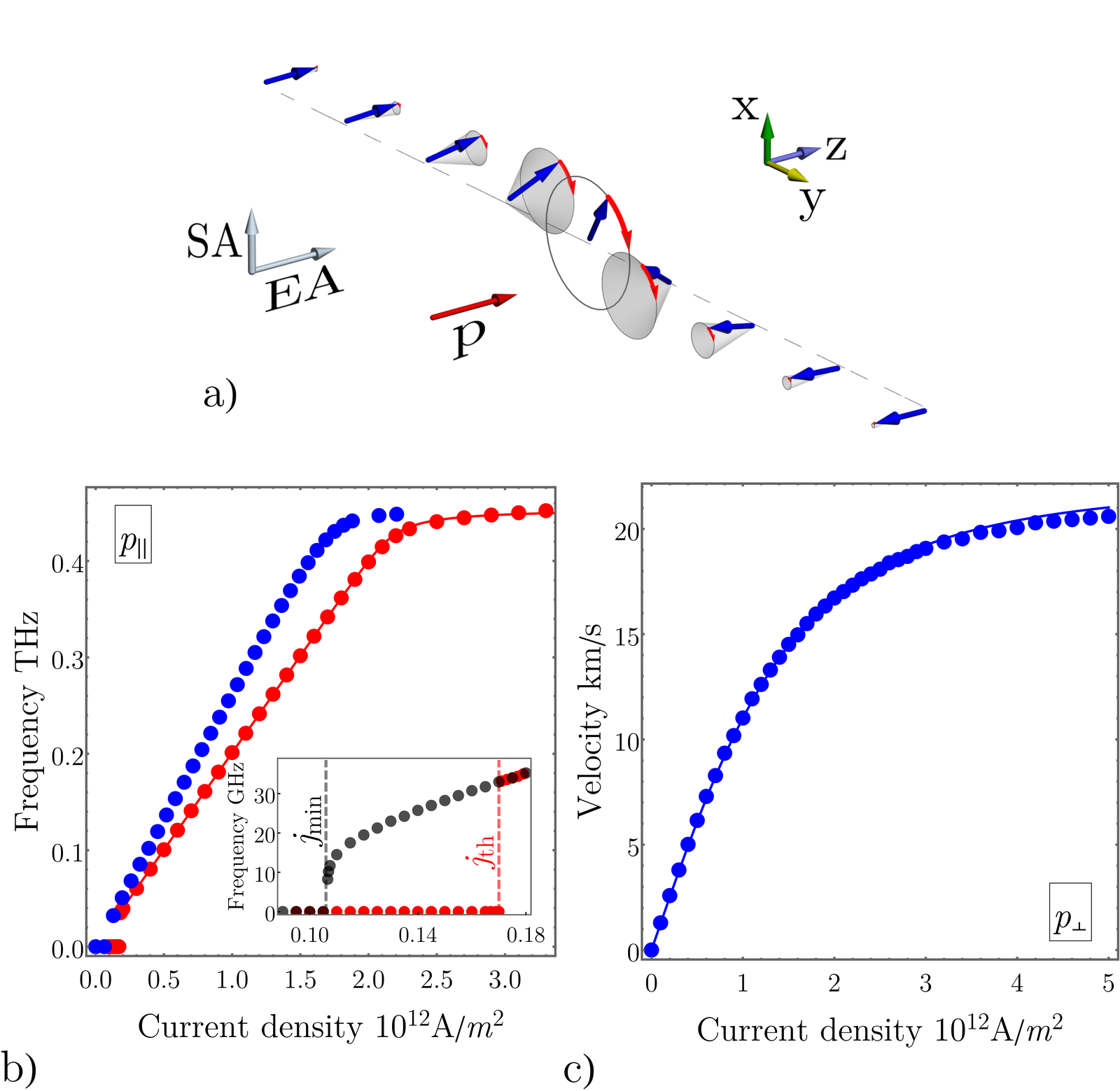}
\caption{\label{fig:b} a) The sketch of a N\'eel vector precession within a DW under the action of direct spin current. b) Frequency $\omega$ of the DW rotation for the NC (blue) and rectangular (red) geometry, c) velocity of the DW as a functions of the dc electric current density. Dots shows the values extracted from the micro-magnetic simulations, while solid lines are calculated analytically. The frequency of the AFM resonance $\omega_0 = 0.45$ THz, the limiting velocity $c = 22$ km/s. }
\end{figure}

 \textit{Parallel polarization:} $\mathbf{p}=\mathbb{z}$, $\mathit{\psi = 0^{\circ}}$.
 The solution for the DW coordinate $X$ corresponds to the standing wall, as $p_x = p_y = 0$ in (\ref{eq:main}). From Eq. (\ref{eq:main2}) follows that spin current,  with polarization along the easy axis $p_z$ excites Josephson-like dynamic \cite{khymyn2017antiferromagnetic} which leads to the rotation of the N\'eel vector within a DW under the action of direct spin current, as it is schematically shown in Fig. \ref{fig:b} a. The precession of the N\'eel vector is of a conical type, similar to the precession of magnetization in ferromagnetic oscillators. However, in ferromagnetic counterparts, the cone opening angle is defined by a balance between applied torque and nonlinear damping. In contrast, here it is determined by a spatial position within a DW, according to the Eq. (\ref{eq:domainWallProfile}) and is independent on torque and damping in a wide frequency range, when $\dot \Phi \ll \omega_{0}$ is negligible in the expression for a DW thickness $\Delta$.
 
Josephson-like spin dynamics, described by Eq. (\ref{eq:main2}) was already widely discussed in the literature in the application to uniform easy-plane AFM materials \cite{khymyn2017antiferromagnetic, khymyn2018ultra}. Eq. (\ref{eq:main2}) is mathematically analogous to the dynamics of a physical pendulum in a gravitational potential under the action of constant external torque. In this analogy, the magnetic anisotropy -- $\omega_{ip}$ plays the role of a gravitational field, and Gilbert damping plays the role of friction. Therefore, the threshold torque (i.e., current) to start auto-oscillations (rotation of the pendulum) is defined only by potential energy at the “top” position, i.e., by the anisotropy value as $\sigma j_{th} = \omega_{ip}^{2}/(2 \omega_{ex})$ and does not depend on damping. The value of $j_{th}$, which is relatively small for our parameters, vanishes for pure uniaxial AFM, and arbitrary weak current excites spin dynamics with a low frequency.
 
In the case of limited spin-current source size with the length $L$, at currents above the threshold, the dependence of the frequency on the current is determined by a balance between the total energy loss in the whole DW and energy gain within the limited spin-current contact area. By integrating the energy balance function with corresponding limits (see Supplemental Materials), one can obtain:
 \begin{equation}
     \alpha \omega = \sigma j \tanh \left( \frac{L}{2 x_0} \sqrt{1 - \frac{\omega^2}{\omega_0^2}} \right).
     \label{eq:frequency}
 \end{equation}
 In the case of a large spin-torque source, $L \gg x_0$, and $\omega \ll \omega_0$, the frequency of the rotation is linearly proportional to the driving current $\omega = \sigma j / \alpha$. The minimum frequency achievable by a constant torque can be obtained by substituting the threshold and reads as $\omega_{th}\simeq \sigma j_{th} / \alpha = \omega_{ip}^2/(2 \alpha \omega_{ex})$ 
 Thus, the Eq. (\ref{eq:frequency}) implies easy tunability of the frequency by the driving current in the range $\omega_{th}..\omega_0$, where $\omega_0$ can reach a sub-THz range. 
 The numerical simulation of such a regime is represented in Fig. \ref{fig:b} b) by symbols, while the solution of analytical Eq. \ref{eq:frequency} for a rectangular sample is shown by a solid red line. The rapid saturation of the generation frequency when approaching the AFM resonance $\omega_0$ is caused by an expansion of the DW thickness $\Delta$ and, hence, by a reduction of a relative overlap with a spin current source. 
 
The first term of Eq. (\ref{eq:main2}), which is inversely proportional to the exchange frequency, also implies an inertial dynamics of the oscillator. Once started, a N\'eel vector will continue to precess even with torques below the aforementioned threshold in the case of a low damping value, which we assume here. To stop oscillations the losses have to overcome the energy gain over the cycle, which gives the minimum current $\sigma j_{min} \simeq 2 \alpha \omega_{ip}/\pi$ of sustained oscillations \cite{khymyn2017antiferromagnetic}, as shown in the inset of Fig. \ref{fig:b} b). To simulate this regime, a short (0.5 ns) rectangular pulse of a current above the threshold ($j>j_{th}$) was applied to start spin dynamics, which was later reduced to the desired values of $j<j_{th}$\cite{khymyn2017antiferromagnetic}. In this way, one can reach arbitrarily low frequencies of the oscillations near $j_{min}$, however, has to take into account a large amplitude of high harmonics \cite{khymyn2018ultra}.
 
Given the Eq. (\ref{eq:outputJ}) for this case, one can notice that the output alternate spin-current is spatially asymmetric:  
\begin{equation}
    \mathbf{\tau}_{out}^y = \omega\sech\left(\frac{x}{\Delta}\right) \tanh\left(\frac{x}{\Delta}\right)\cos \omega t
    \label{eq:output1}
\end{equation}

\begin{figure}[b!]
\includegraphics[width=\linewidth]{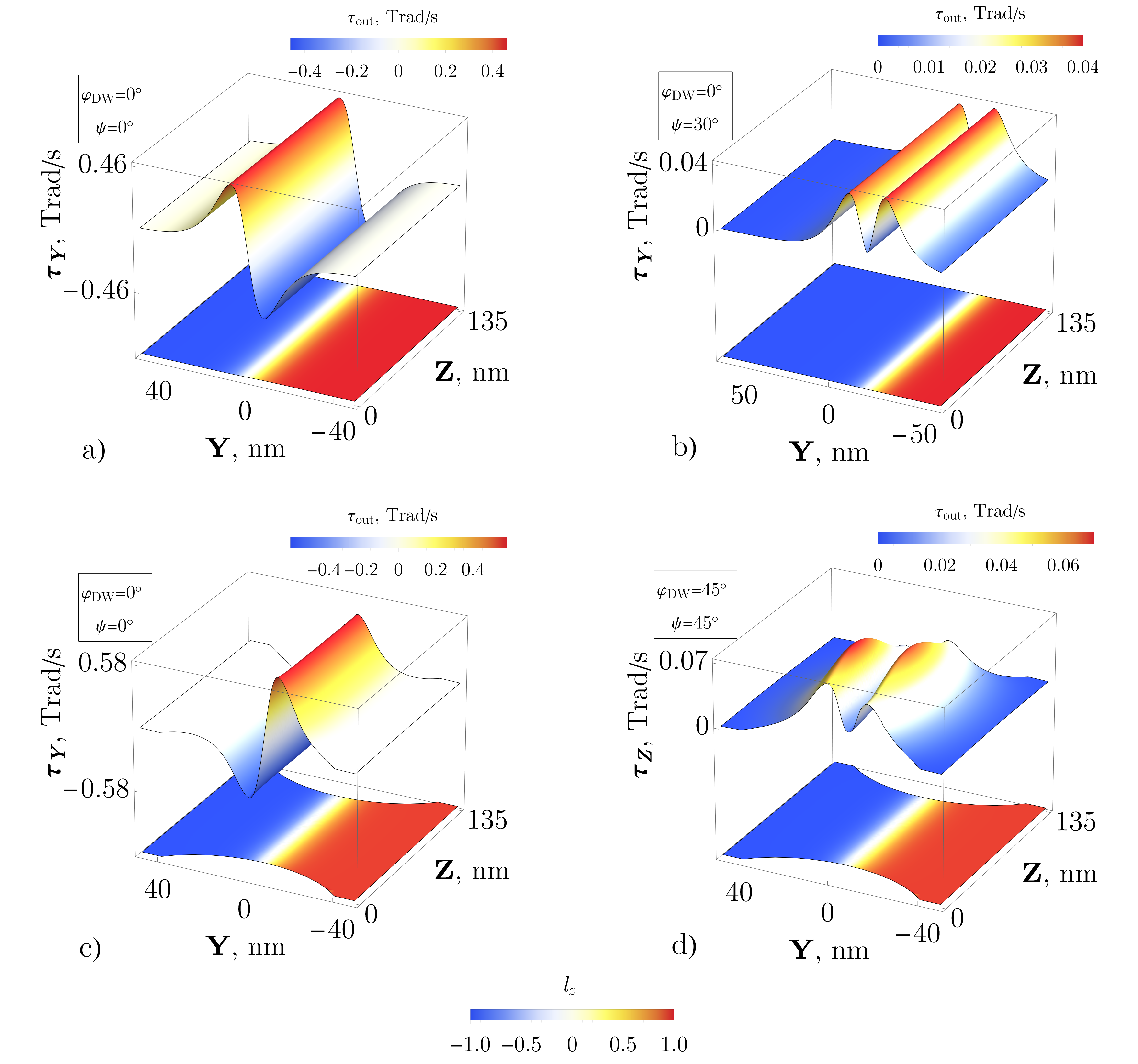}
\caption{\label{fig:outputTorque} Simulated distribution of the x-component of the output torque $\mathbf{\tau}_{out} = \mathbf{l} \times \partial \mathbf{l} / \partial t$ (the upper plane in each figure) and $z$-component of the Neel vector (the lower one) in the plane of the film for different geometries: rectangular a), b) and NC c), d); and different spin polarization angle: $\psi = 0^{\circ}$ a), c), $\psi = 30^{\circ}$ b) and $\psi = 45^{\circ}$ d) . The applied current density is $7.8 \cdot 10^{11}$ A/m$^2$. (a) and (c) show alternate torque at the primary frequency of oscillations $\omega/2\pi=160$ GHz (Eq. \ref{eq:output1}), while (b) and (d) at the doubled frequency $2\omega/2\pi=320$ GHz (Eq. \ref{eq:output2}).} 
\end{figure}

Since $\mathbf{I}_{out} \propto \int \mathbf{\tau}_{out} dx dy$ averaged over the DW vanishes, see Fig. \ref{fig:outputTorque} (a, b), the signals from the film areas, where $\sin \theta \cos \theta \lessgtr 0$ have to be read out independently that can be achieved by an additional spin-Hall electrode on top of the AFM layer. As well, a unique technique used in work \cite{li2020spin} can be applied, where two nanowires made of metals with opposite signs of the spin-Hall angle (such as e.g., tantalum and platinum) are placed to the opposite slopes of the DW to sum up the signals (see, Fig. \ref{fig:schema}b). 

\textit{Perpendicular polarization:} $\mathbf{p}=\mathbb{y}$, $\mathit{\psi = 90^{\circ}}$.
In this limit case, the precessional dynamics is not excited, and only the translational motion of the DW is possible that was already highlighted in literature \cite{shiino2016antiferromagnetic, sanchez2020dynamics, hals2011phenomenology}. The equation for the velocity of a DW motion is common to that for a DW driven by the N\'eel spin-orbit torques, considered in details in  Ref. \cite{gomonay2016high}(see also Supplemental Materials for details), namely:
\begin{equation}
    v = \left. \mu \tau c\middle/\sqrt{(\mu \tau)^2 + c^2}\right.,
\end{equation}
where $\mu = \pi x_0 / 2\alpha$ has the sense of the mobility of the domain wall ($v\simeq \mu \tau$ at $\mu \tau \ll c$). The dependence of the velocity on the applied current for the rectangular geometry is shown on Fig. \ref{fig:b} c).

\textit{Oblique polarization} $\mathit{0^{\circ} < \psi < 90^{\circ}}$. In this case, the torque component ($p_z$) defines the solution for the angle $\Phi$ in Eq. (\ref{eq:main2}) and, thus, controls the efficiency of the driving force in Eq. (\ref{eq:main}).
As in the case of $\psi=0$, the anisotropy in the hard plane induces the threshold for the oscillations. 
Above the threshold $j \gg j_{th}$, particularly for pure uniaxial AFM, when $j_{th}=0$, the solution takes the  form $\Phi \approx \omega t$ that leads to oscillations of the domain wall around the equilibrium position with the velocity defined by:
\begin{equation}
   \frac{\dot X}{\sqrt{1 - \dot X^2 / c^2}} = \Delta(\omega) \frac{\pi p_y}{2} \frac{\tau \omega_{ex}}{\sqrt{\omega^2 +  \alpha^2\omega_{ex}^2}}\sin(\omega t + \varphi),
\end{equation}
where  $\Delta(\omega) = x_0 \left. \middle/\sqrt{1 - \omega^2/\omega_0^2}\right.$ and $\varphi = \arctan (\alpha \omega_{ex}/\omega)$. 

Taking for simplicity $\dot X \ll c$, the coordinate can be written as:
\begin{equation}\label{eq:XvsT}
    X(t) = X_{max} \cos ( \omega t + \varphi ),
\end{equation}
where the amplitude of the DW translational oscillations is defined by:
\begin{equation}
    X_{max} =\frac{\pi p_y}{2 p_z} \frac{\alpha \omega_{ex}}{   \sqrt{\omega^{2} + \alpha^{2} \omega_{ex}^{2} } } \Delta(\omega)
\end{equation}

\begin{figure}[hbt!]
\includegraphics[width=0.95\linewidth]{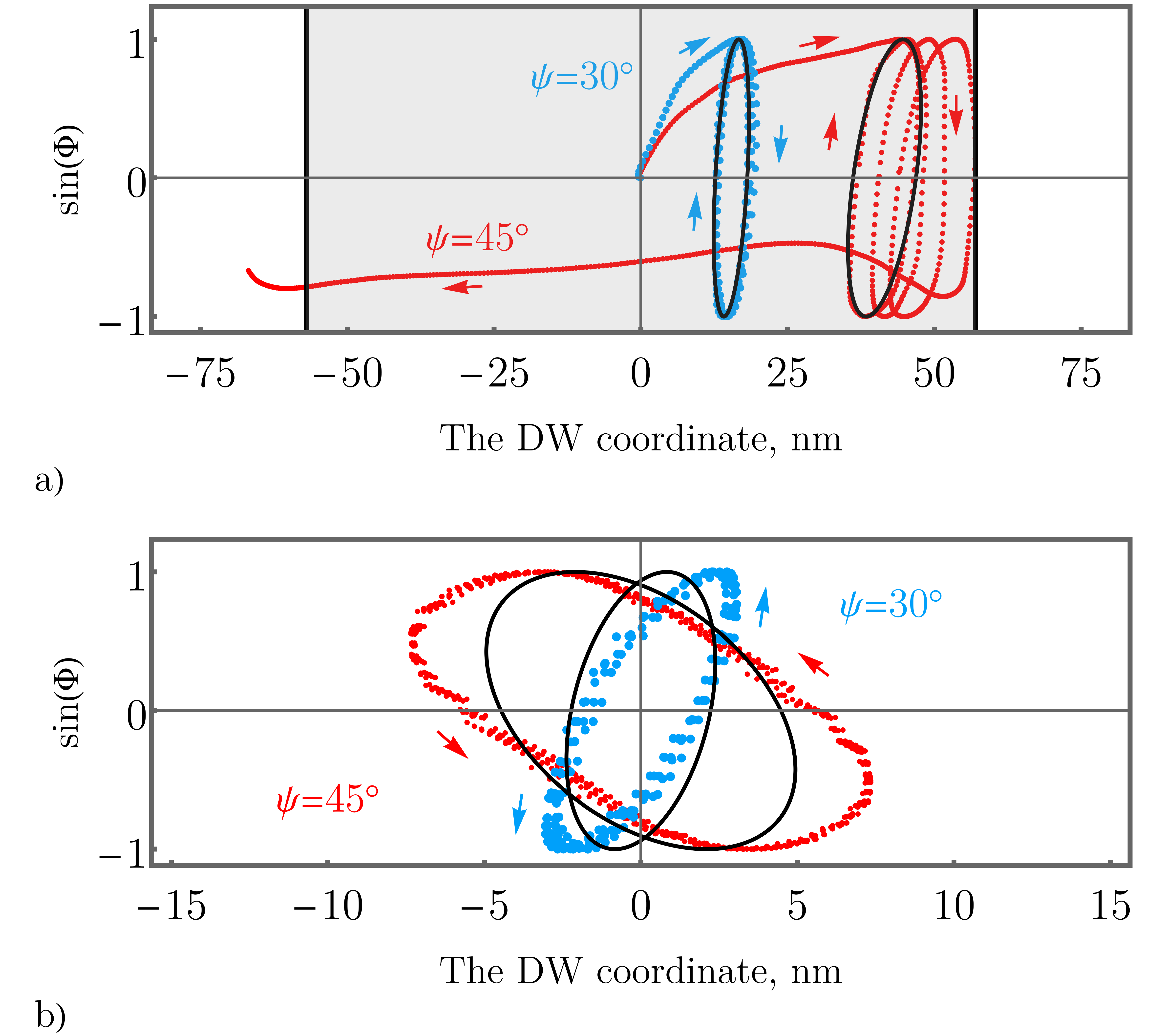}
\caption{\label{fig:phiVsX} The phase of the domain wall $\Phi$ as a function of the domain wall coordinate $X$ for a) the rectangular and b) the NC geometries. Grey rectangle defines the spin-Hall region, were current is flowing; black ellipses show analytical dependence (\ref{eq:XvsT}). The applied current density is $2.4 \cdot 10^{11}$ A/m$^2$. }
\end{figure}

However, the analytical solution $\Phi = \omega t$ is approximate \cite{khymyn2017antiferromagnetic}, and our micro-magnetic simulations show that periodic wall oscillations are accompanied by a drift, see Fig. \ref{fig:phiVsX} a). This drift is a significant problem for the signal readout in a rectangular geometry; depending on the polarization of the spin current, angle $\Phi$, and velocity of the domain wall, it can bounce off the edge of the contact, stay on edge for a long time, or go beyond the STT source area, which will disrupt signal generation (see Fig. \ref{fig:phiVsX} a) for $\psi=45^\circ$). This problem can be solved by pinning the wall, such as using NC, see Fig. \ref{fig:phiVsX} b). The NC creates a restoring force for the DW, so as can be seen from the simulations, the wall ceases to drift, and the points of the phase change are fixed in coordinate. Interestingly, the amplitude of the oscillations increases as the DW exhibits resonant behavior in the pinning potential.

The translational oscillations of the DW produce an additional output torque at the doubled frequency:
\begin{equation}\label{eq:output2}
       \mathbf{\tau}_{out}^{y} = \frac{\omega X_{max}}{\Delta} \sech\left(\frac{x - X(t)}{\Delta}\right) \cos 2 \omega t
\end{equation}

In contrast to Eq. (\ref{eq:output1}), the signal is symmetrical in coordinate, see Fig. \ref{fig:outputTorque} (c, d), so there is no need for the additional readout layers. However, $\mathbf{\tau}_{out}^{y}$ creates an electrical current in $z$-direction, i.e., perpendicular to the input one, which requires a complex four-terminal device design. To avoid this issue, one can tilt the easy axis of the AFM in the $yz$-plane by an angle $\varphi_{DW}$, and as a result, an alternating $z$-component appears in the output torque. In this case, the same two terminals can be used simultaneously as a source of the dc input and a detector of the ac output as in conventional spin-Hall oscillators. 

\begin{figure}[hbt!]
\includegraphics[width=0.95\linewidth]{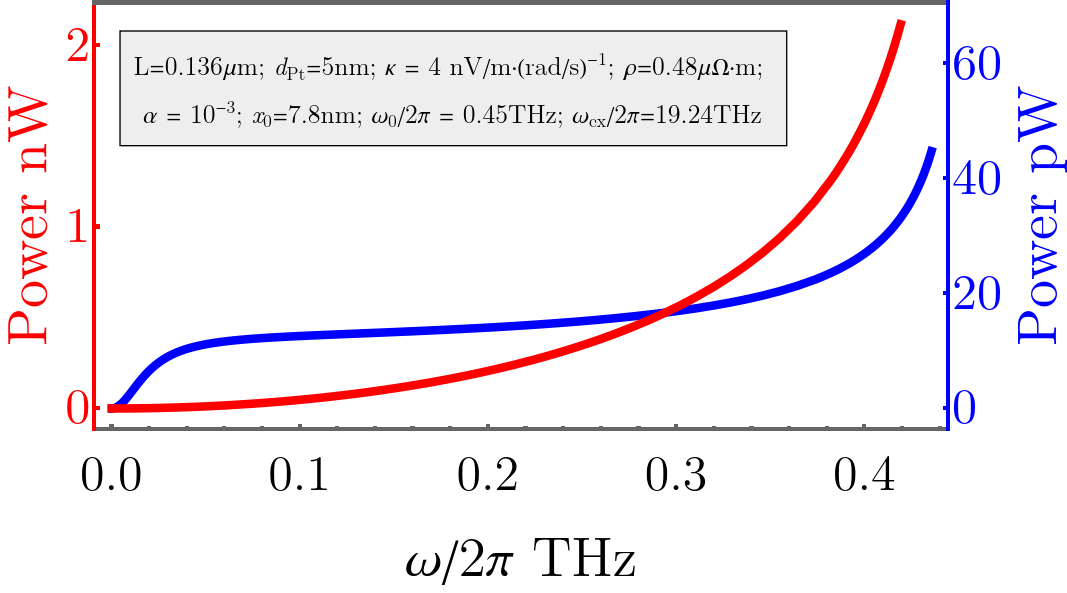}
\caption{\label{fig:power} The output power of the spin-Hall nanooscillator. The red line corresponds to $\psi = 0^{\circ}, \phi_{DW} = 0^{\circ}$ and is extracted at the main frequency of oscillations $\omega$, while the blue line - at the doubled $2\omega$ and $\psi = 45^{\circ}, \phi_{DW} = 45^{\circ}$.}
\end{figure}

The output power of the SHNO can be calculated using the method described in  Ref. \cite{khymyn2017antiferromagnetic}. It is presented in Fig. \ref{fig:power} for the aforementioned parameters of the SHNO. The useful signal at the frequency $\omega$ of the DW precession achieves higher power, although it requires additional readout layers. However, even at the doubled frequency $2\omega$ and simple bilayer SHNO layout, the output power can reach tens of picowatts for $100$ nm DW length, which is a few times higher than for the conventional ferromagnetic devices \cite{mohseni2013spin}. It is noticeable that a power delivered by both output methods substantially grows near the AFM resonance frequency $\omega_0$, which makes it preferable to operate in the high-frequency range.

In conclusion, we showed that the spin current can excite complex dynamical regimes for an AFM DW, which include the precession of the Neel vector in the center of a DW, as well as oscillatory motion of the DW position. This dynamics provides a substantial alternate output spin current, which can be used for the development of efficient subterahertz spin-Hall nano-oscillators. 
The pronounced feature of this oscillator is the possibility to minimize a threshold current till complete vanishing for a pure uniaxial case and the increasing of the output power with the frequency, see Fig. \ref{fig:power}. 

This project has received funding from the European Research Council
(ERC) under the European Union’s Horizon 2020 research and innovation programme (grant agreement TOPSPIN No 835068). B. A. Ivanov acknowledges support by National Research Foundation of Ukraine, Grant No. 2020.02/0261.

\nocite{*}
\bibliography{ArxivKhymynAFMDWOscillator}

\end{document}


\maketitle

The Lagrangian of the purely  uniaxial antiferromagnet can be represented in the form:
 \begin{equation}
 \mathcal{L}=\frac{M_s}{2 \gamma \omega_{ex}}\left[\dot \theta^2 -c^2 \theta'^2+\sin^2 \theta (\dot \phi^2 -c^2 \phi'^2 - \omega_0^2)\right]
 \label{lagrangian}
 \end{equation}
The rate of the energy losses due to the dissipation and the action of spin-transfer torque can be written as:
\begin{align} 
\dot E  = - \dot{\mathbf{l}} \cdot \frac{\partial \mathcal{R}}{\partial \dot{\mathbf{l}}} &=-\frac{ \alpha M_s }{\gamma}(\dot \theta^2+\dot\phi^2 \sin^2 \theta )- \nonumber\\ 
 &-\frac{\tau M_s }{\gamma}[p_x(\dot\theta\sin\phi +\dot\phi\cos\phi \sin 2 \theta) - p_y(\dot\theta\cos\phi -\dot\phi\sin\phi \sin 2 \theta) -p_z\dot\phi\sin^2 \theta]
 \label{dissip}
\end{align}

The forced dynamics of a domain wall under the action of such perturbations can be described by the two collective coordinates, the displacement $X$ and the phase $\Phi$ of the domain  wall. This approach corresponds to the so-called adiabatic approximation in perturbation theory for solitons\cite{kivshar1989dynamics}, which can be applied if the characteristic perturbation parameters are small. For instance, if the effective dissipation constant $\alpha \ll 1$, the amplitude of the spin current (in the units of frequency) $\tau \sim \alpha\omega_0 $ and the rates of change of the domain wall velocity $v = \dot X$ and its precession frequency $\omega = \dot \Phi$ are of the order of small system parameters $dv / dt \sim \alpha \omega_0 v \ll \omega_0 v$, $d\omega / dt \sim \alpha \omega_0 \omega \ll \omega_0 \omega$. The conditions $\alpha \ll 1$ and $\tau \sim \alpha \omega_0 $ will be considered as being fulfilled as they are satisfied for magnets that are interesting for spintronics applications. The exact parametrized solution of the domain wall\cite{galkina2017precessional, kim2014propulsion} is:
\begin{equation}
\cos \theta=\tanh \left(\frac{x-X(t)}{\Delta}\right),  \qquad \phi= \Phi(t),
\label{domainWallProfile}
\end{equation}
where $\Delta$ means the DW thickness, with the next ansatz:
\begin{equation}
    \Delta = x_0 \frac{\sqrt{1 - \frac{\dot X^2}{c^2}}}{\sqrt{1 - \frac{\dot \Phi^2}{\omega_0^2}}}
\end{equation}

Substituting (\ref{domainWallProfile}) into (\ref{lagrangian}) and (\ref{dissip}) and integrating  over the $x$, we obtain the effective Lagrangian and  energy losses functions of the dynamic system with two degrees of freedom, which describe the dynamics of the collective coordinates:

\begin{equation}
\mathcal{L}= - \frac{2 M_s}{\gamma \omega_{ex}} c \omega_0 \sqrt{1 - \frac{\dot{X}^2}{c^2}} \sqrt{1 - \frac{\dot \Phi^2}{ \omega_0^2 } } \label{lagrangianColCoord},
\end{equation}

\begin{eqnarray}
\dot E = -\frac{2 \alpha M_s }{\gamma}\left(\frac{\dot{X}^2}{\Delta}     +   \Delta\dot\Phi^2  \right) - \frac{ \tau M_s }{\gamma}\left[\pi \dot{X}(p_x  \sin\Phi - p_y   \cos\Phi) -p_z 2\Delta \dot\Phi \right]\label{eq:edot}
\end{eqnarray}

The Lagrangian (\ref{lagrangianColCoord}) describes a free relativistic particle. The energy losses can be represented as the sum of two independent terms, products of linear and angular velocities with corresponding friction forces:

\begin{eqnarray}
\dot E= \dot X F_X + \dot \Phi F_{\Phi} =\dot X \left(   -\frac{2 \alpha M_s }{\gamma}\frac{\dot{X}}{\Delta}   - \frac{ \pi \tau M_s }{\gamma}(p_x  \sin\Phi - p_y   \cos\Phi)  \right) + \dot \Phi \left(
 -\frac{2\alpha M_s }{\gamma}\Delta \dot \Phi +  \frac{2\tau M_s }{\gamma} p_z \Delta
\right)
\end{eqnarray}

Therefore, we can build the relativistic version of Newton’s second law for collective coordinates, the time derivative of generalized momentum $P_{q} = \partial \mathcal{L}(q, \dot q, t)/\partial  \dot q$ (where $q = X, \Phi$) is equal to the corresponding force acting on the particle:
\begin{eqnarray}
    &&\frac{1}{\omega_{ex}} \frac{d}{dt} \left( \frac{ \dot X }{ \Delta} \right)  +   \alpha \frac{\dot{X}}{\Delta}  +  \tau\frac{\pi }{2}( p_x \sin \Phi - p_y  \cos \Phi) = 0, \\
       &&\frac{1}{ \omega_{ex}} \frac{d}{dt} ( \Delta \dot \Phi) + \alpha \Delta \dot \Phi  - \Delta\tau p_z   + \Delta\frac{\omega_{ip}^2}{2\omega_{ex}} \sin 2\Phi = 0,
\end{eqnarray}
where we additionally took into account the presence of a small anisotropy in the base plane $\omega_{in} \ll \omega_0$. For example, these equations have two particular solutions. The first one corresponds to a steady-state motion of the domain wall\cite{shiino2016antiferromagnetic, gomonay2016high} $X = vt$ if $p_z = 0$. The velocity of this motion for the case $\Phi = 0$ is defined as:

\begin{equation}
    v = \frac{\mu \tau c}{ \sqrt{(\mu \tau)^2 + c^2}} ,
\end{equation}
where $\mu = p_y \pi x_0 / 2\alpha$ has the sense of the mobility of the domain wall with respect to $\tau$, $ v = \mu \tau$ at a low velocity. The second solution, for the case $p_x,p_y = 0$ and $\omega_{ip} = 0$, describes a uniform in time precession $\Phi = \omega t$ with frequency:

\begin{equation}
    \omega = \tau p_z / \alpha\label{solutionPhiSimple}
\end{equation}

\textcolor{blue}{
Here we should note that the effect of saturation of the precession frequency in the domain wall, discussed in the paper, is associated with a limited length of the spin current source. In the case of a limited spin current  contact length, the frequency dependence can be found from the power balance regime $\dot{E} = 0$ by substituting $\sin \theta = \sech \left[x/\Delta(\omega)\right]$ in (\ref{dissip})  and integrating the dissipative part (i.e. the first term in (\ref{dissip})) over the whole DW ($x \in (-\infty,\infty)$) and STT part over the spin current source area ($x \in (-L/2, L/2))$:  }
\begin{equation}
    \int_{-\infty}^{\infty} \alpha \omega \sin^2 \theta dx = \int_{- \frac{L}{2}}^{\frac{L}{2}} \tau p_z \sin^2 \theta dx,
\end{equation}
which leads to the equation for the frequency:
\begin{equation}
     \alpha \omega = \sigma j \tanh \left( \frac{L}{2 x_0} \sqrt{1 - \frac{\omega^2}{\omega_0^2}} \right).
     \label{eq:frequency}
\end{equation}
\textcolor{blue}{As the frequency increases and approaches $\omega_0$, the wall expands and its size $\Delta$ may become larger than the size of the spin current source. }

The presence of in-plane anisotropy creates a treshold for excitation $\tau^{th}p_z = \omega_{ip}^2/2\omega_{ex}$ and leads to a non-uniform in time precession dynamics, an approximate analytical solution for supercritical torques $\tau \gg \tau^{th}$\cite{khymyn2017antiferromagnetic} is given by:

\begin{equation}
    \dot\Phi(t) \approx \omega + \frac{\omega_{ip}^2}{4 \sqrt{\alpha^2 \omega_{ex}^2 + 4\omega^2}}\cos 2 \omega t,
\end{equation}
where $\omega$ is defined by (\ref{solutionPhiSimple}).

\printbibliography